\documentclass[10pt,letterpaper]{article}
\usepackage{opex3}
\usepackage{color}

\begin{document}

%%%%%%%%%%%%%%%%%% title page information %%%%%%%%%%%%%%%%%%
\title{Photophysics of  single silicon vacancy centers in diamond: implications for single photon emission}

\author{Elke Neu$^1$, Mario Agio$^2$, and Christoph Becher$^{1,\star}$}
\address{${}^1$Universit\"at des Saarlandes, Fachrichtung 7.2 (Experimentalphysik), 66123 Saarbr\"ucken, Germany

${}^2$National Institute of Optics (INO-CNR) and European Laboratory for Nonlinear Spectroscopy (LENS), 50019 Sesto Fiorentino, Italy}

\email{$^{\star}$christoph.becher@physik.uni-saarland.de} %% email address is required

% \homepage{http:...} %% author's URL, if desired

%%%%%%%%%%%%%%%%%%% abstract and OCIS codes %%%%%%%%%%%%%%%%
%% [use \begin{abstract*}...\end{abstract*} if exempt from copyright]

\begin{abstract}
Single silicon vacancy (SiV) color centers in diamond have recently shown the ability for high brightness, narrow bandwidth, room temperature single photon emission. This work develops a model describing the three level population dynamics of single SiV centers in diamond nanocrystals on iridium surfaces including an intensity dependent de-shelving process. Furthermore, we investigate the brightness and photostability of single centers and find maximum single photon rates of 6.2 Mcps under continuous excitation. We investigate the collection efficiency of the fluorescence and estimate quantum efficiencies of the SiV centers.
\end{abstract}

 % REPLACE WITH CORRECT OCIS CODES FOR YOUR ARTICLE

%%%%%%%%%%%%%%%%%%%%%%% References %%%%%%%%%%%%%%%%%%%%%%%%%

%%%%%%%%%%%%%%%%%%%%%%%%%%  body  %%%%%%%%%%%%%%%%%%%%%%%%%%
\section{Introduction}
Single color centers in diamond are auspicious for applications as solid state single photon sources (for a review see \cite{Aharonovich2011}). Silicon vacancy (SiV) centers are especially promising due to their spectral properties such as  the concentration of the fluorescence in a narrow zero-phonon-line (ZPL) with a room temperature width of down to 0.7 nm \cite{Neu2011}. Furthermore, their emission is situated in the red spectral region at 740 nm, in a wavelength range where the background fluorescence of the diamond material is low \cite{Neu2011a}. Studies using single SiV centers created via ion implantation in natural diamond, however, revealed a low brightness (approx. 1000 cps) \cite{Wang2006,Wang2007b}. More recently, bright single SiV centers with up to 4.8 Mcps, created \textit{in situ}, i.e., during the chemical vapor deposition (CVD) growth of randomly oriented nanodiamonds (NDs) \cite{Neu2011} and (001) oriented heteroepitaxial nanoislands (NIs) \cite{Neu2011b} on iridium (Ir) films, have been observed. The origin of the enhanced brightness has not been fully explored up to now. In this context, it is crucial to investigate the collection efficiency obtained in the system as well as the quantum efficiency of the SiV centers.

In this paper, we analyze several bright SiV centers in detail. The investigated SiV centers are hosted by NDs or NIs on Ir as introduced above (for sample details see \cite{Neu2011,Neu2011b}). Note that we additionally use a ND sample that has been grown with slightly modified CVD parameters compared to \cite{Neu2011} (55 min growth duration, 0.4\% CH$_4$) and contains slightly larger nanodiamonds (220 nm mean size). The SiV centers in this slightly modified sample showed similar characteristics. We extensively investigate the population dynamics of 14 single SiV centers including the saturation of the photoluminescence and explore the underlying level scheme, verifying a model including intensity dependent de-shelving which we suggested in \cite{Neu2011} based only on the analysis of the population dynamics of one single SiV center. Furthermore, we characterize the photostability of single SiV centers. We calculate the collection efficiency for the fluorescence of SiV centers on Ir. From this calculation and the maximum excited state population, we estimate the quantum efficiency for the ZPL transition and indicate possible origins of non-radiative decay.
\section{Brightness of single SiV centers}\label{sec_satroomtemp}
\begin{figure}[h]
\centering
\includegraphics[width=0.95\textwidth]{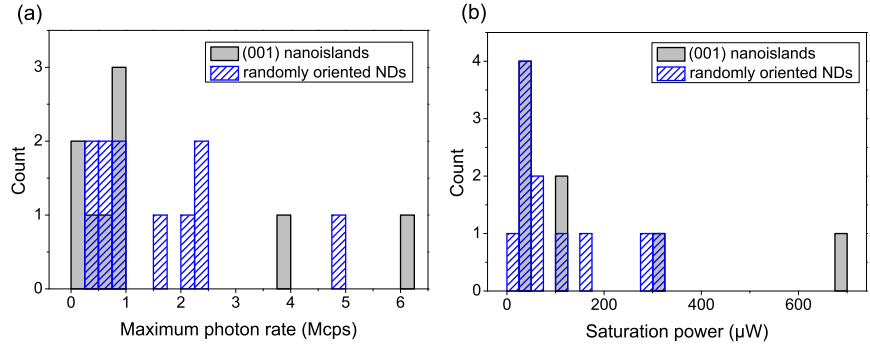}
\caption{Histograms of (a) maximum obtainable photon rates at saturation $I_{\infty}$ (b) saturation powers $P_{sat}$. Histograms take into account emitters from randomly oriented NDs and (001) NIs.  NDs: excitation at 671 nm. NIs:  excitation at 695--696 nm.\label{sathist}}
\end{figure}

As an important figure of merit for a single photon source, we first determine the maximum photon count rate $I_\infty$. For this experiment, we use a confocal microscope setup which is described in detail in \cite{Neu2011,Neu2011b}. For the randomly oriented NDs, excitation at 671 nm was employed; for the (001) NIs, an excitation wavelength of 695--696 nm was used (unless otherwise stated).
As single SiV centers show preferential absorption of linearly polarized light \cite{Neu2011b}, we employ the optimized linear polarization direction for excitation. All count rates are corrected for the dark counts of the setup. In the presence of background fluorescence, the fluorescence rate $I$ of a single emitter as a function of the excitation power $P$ is described by
\begin{equation}
      I=I_\infty\frac{P}{P+P_{sat}}+c_{backgr}P.
      \label{Satfunc2}
\end{equation}
Using Eq.\ (\ref{Satfunc2}), we fit the saturation curves for single SiV centers and obtain the saturation powers $P_{sat}$ and maximum photon rates $I_{\infty}$ summarized in Fig.\ \ref{sathist}. We find a mean value for $I_{\infty}$ of $1.5\pm1.4$ Mcps in the randomly oriented NDs and of $1.5\pm2$ Mcps in the (001) oriented NIs. The high standard deviation illustrates the variation in brightness of the emitters which will be discussed in detail in Secs.\ \ref{g2threelevel} and \ref{sec_quantumeff_colleff}. The highest $I_{\infty}$ obtained from the fits is 6.2 Mcps. Thus, the single emitters observed here are the brightest color centers to date under continuous laser excitation.

As apparent from Fig.\ \ref{sathist}(b), also $P_{sat}$ displays a significant spread among different emitters. The highest value observed is 692 $\mu$W, the lowest 14.3 $\mu$W with a mean value of $105\pm103$ $\mu$W for the NDs and  $175\pm229$ $\mu$W for the NIs. By taking into account the transmission of the laser light through the microscope objective and the spot size of the focus (1/e$^2$ radius) of approx. 0.5 $\mu$m, we estimate the intensity maximum of the focused beam impinging onto the color centers. The highest and lowest values of $P_{sat}$ correspond to an intensity of 58.8 kW/cm$^2$ and 1.2 kW/cm$^2$. Thus, 2.1$\times$10$^{23}$ photons$\cdot$s$^{-1}\cdot$cm$^{-2}$ impinge onto the emitter with the highest $P_{sat}$ (excitation at 695 nm), whereas only 4.1$\times$10$^{21}$ photons$\cdot$s$^{-1}\cdot$cm$^{-2}$ are present for the emitter with the lowest $P_{sat}$ (excitation at 671 nm). 1 $\mu$W at a wavelength of 671 nm (695 nm) corresponds to 2.87 (2.97)$\times$10$^{20}$ photons$\cdot$s$^{-1}\cdot$cm$^{-2}$. Note that the estimation of the intensity does not take into account the iridium surface, any losses (reflection, scattering, absorption) due to the nanodiamond or the occurrence of localized modes in the nanocrystals (discussion see below). The observed values of $P_{sat}$ are significantly smaller compared to previous studies using single SiV centers implanted into natural bulk diamonds: Wang et al. \cite{Wang2006} report $P_{sat}=6.9$ mW (excitation 685 nm, comparable or even tighter focussing of the excitation laser). The more efficient excitation, first, might arise from local field enhancements at the site of the SiV centers: In (spherical) NDs with sizes comparable to the wavelength of the excitation/fluorescence light, resonant modes (Mie resonances) of the light field can develop \cite{Greffet2011}. The excitation laser light is coupled into these modes; the resulting field distribution excites the color center \cite{Greffet2011}. Depending on the position of the color center, it experiences a high or low excitation light intensity. The enhanced or reduced intensity, compared to the situation in bulk diamond where the Gaussian focus of the laser determines the local intensity, leads to a lower or higher  $P_{sat}$. This effect is indistinguishable from an altered absorption coefficient as it results in a change of $P_{sat}$ as well. Note that as the NDs/NIs used here are not spherical, the formalism used in Ref.\ \cite{Greffet2011} cannot be straightforwardly applied. Selecting emitters from a confocal fluorescence map introduces an experimental bias: one preferably selects bright single centers most probably experiencing high local intensity and thus efficient excitation. Thus, the histograms of Fig.\ \ref{sathist} tend to summarize SiV centers which experience efficient excitation despite the fact that the sample might contain centers where local fields introduce less efficient excitation.

\label{absdisc5}A second cause for a reduced $P_{sat}$  might be an enhanced absorption coefficient. The ground state of the SiV center was reported about 2.05 eV below the conduction band edge \cite{Iakoubovskii2000b}, excluding an excitation of the color center's electrons into the conduction band for our 1.78 eV (695 nm) or 1.85 eV (671 nm) excitation. The excitation, therefore, most probably involves excited vibrational states of the electronically excited state. Individual SiV centers show strongly varying vibronic sideband spectra in emission together with a varying overall emission into the sidebands \cite{Neu2011b}. Changes in the emission spectrum can also indicate changes in the absorption spectrum \cite{Osadko1979}, thus potentially altering the absorption coefficient for a given excitation wavelength. In previous work by Wang \cite{Wang2006,Wang2007b}, no sideband spectra were recorded precluding direct comparison.
\section{Intensity auto-correlation ($g^{(2)}$) measurements} \label{g2threelevel}
\begin{figure}[t]
\centering
\includegraphics[width=0.9\textwidth]{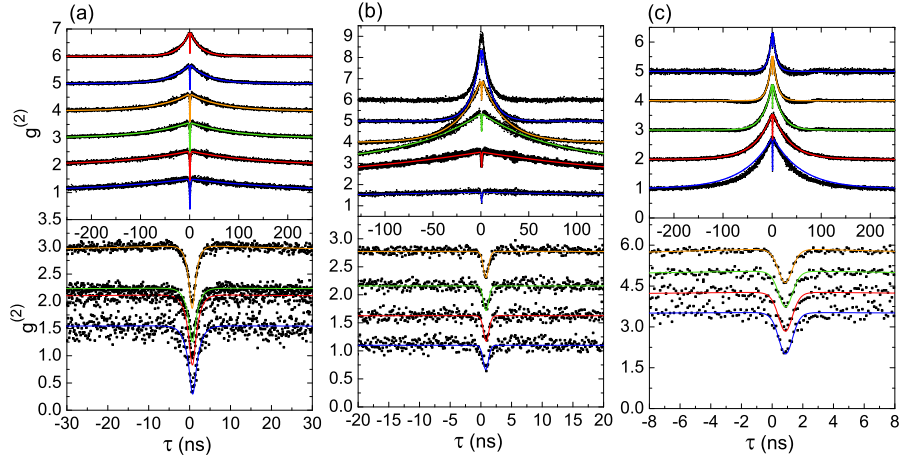}
\caption{$g^{(2)}$ functions for varying excitation power. Consecutive $g^{(2)}$ functions have been shifted for clarity (lowest excitation power $g^{(2)}$ function in each graph not shifted, higher (lower) excitation power $g^{(2)}$ functions shifted by 1 (0.5) each). (a) emitter ND3 (lower graph $0.03,0.11,0.15,0.49 P_{sat}$, upper graph $0.6,0.8,1.2,1.7,3.5,5.8 P_{sat}$). (b) emitter ND1, $g^{(2)}$ includes background correction  (lower graph $0.08,0.17,0.28,0.47 P_{sat}$, upper graph $1.5,3.3,6.3,11.9,23.5,32.7 P_{sat}$). (c) emitter NI1(lower graph $0.01,0.02,0.03,0.07 P_{sat}$, upper graph $0.2,0.3,0.6,1.1,1.4 P_{sat}$).     \label{fig:g2examples} }
\end{figure}

\begin{figure}[h]
\centering
\includegraphics[width=0.35\textwidth]{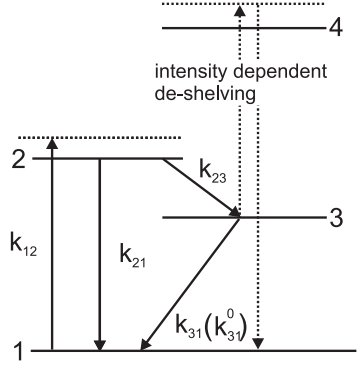}
\caption{Schematic representation of the extended three level model employed to explain the population dynamics of single SiV centers, explanation see text. \label{3levelschemedesh}}
\end{figure}

\begin{figure}
\centering
\includegraphics[width=0.9\textwidth]{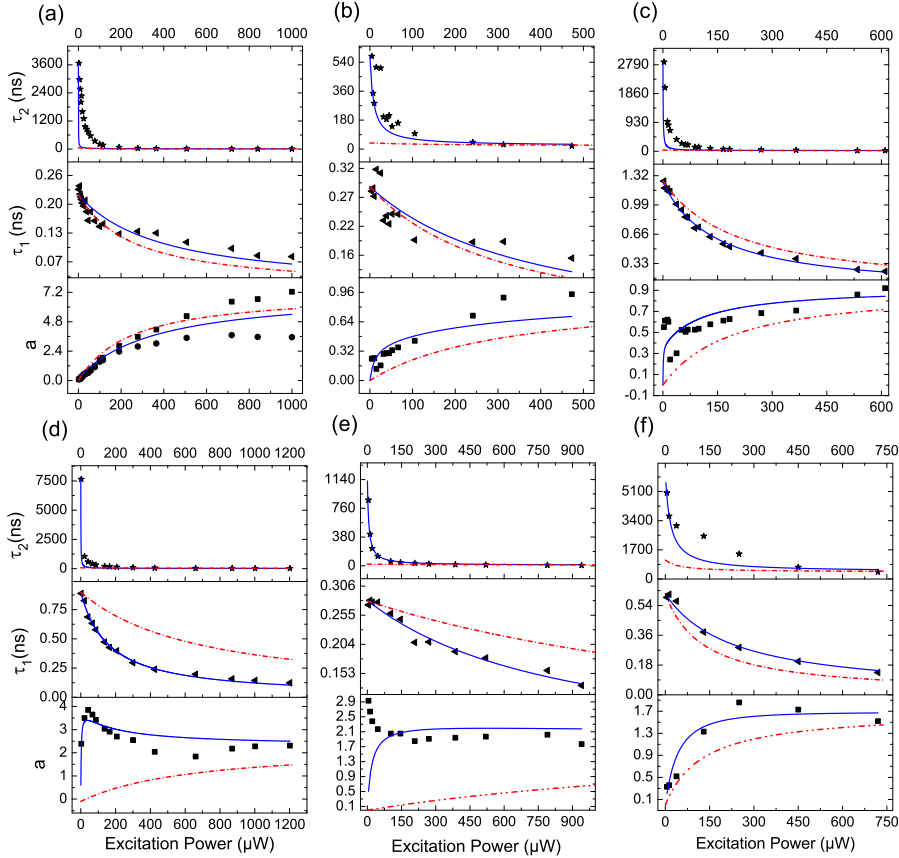}
\caption{Intensity dependence of the parameters $a$, $\tau_1$ and $\tau_2$: (a) emitter ND1, (b) ND2, (c) ND3, (d) ND4 (exc. 705 nm), (e) NI1, (f) NI7. Blue solid lines represent fitting curves according to the intensity dependent de-shelving model. Red dashed lines represent the excitation power dependence from the model with constant rates.  \label{g2par} }
\end{figure}

Figure \ref{fig:g2examples} exemplarily displays excitation power dependent $g^{(2)}$ functions for three individual SiV centers. The $g^{(2)}$ functions have been normalized assuming that $g^{(2)}(\tau)=1$ for long delay times $\tau$. All measurements reveal a distinct antibunching. Furthermore, the $g^{(2)}$ functions exceed one for certain delay times (bunching). NI labels emitters located in nanoislands, ND labels emitters located in randomly oriented nanodiamonds throughout this work. For emitter NI1, a pronounced bunching already occurs at low excitation powers, while for emitter ND1 it only becomes visible at elevated excitation powers. $g^{(2)}$ functions involving a bunching indicate a three level system. In a first approach, we use a simplified model depicted in Fig.\ \ref{3levelschemedesh} for the population dynamics: Levels 1 and 2 are coupled via a fast radiative transition (rate coefficient $k_{21}$), the photons emitted on this transition are detected to determine $g^{(2)}$. In contrast, level 3 acts as a shelving state populated via the rate coefficient $k_{23}$ with the possibility of relaxation into the ground state via $k_{31}$. As long as the emitter resides in state 3, no photons on the transition 2$\rightarrow$1 are detected. This simple model has been successfully applied to molecules involving shelving states \cite{Kitson1998}. To obtain the $g^{(2)}$ function, one solves the rate equations for the populations $N_i$ resulting in
\begin{equation}
g^{(2)}(\tau)=1-(1+a)\,e^{-|\tau|/\tau_1}+a\,e^{-|\tau|/\tau_2}
\label{g23level}
\end{equation}
The parameters $a$, $\tau_1$ and $\tau_2$ are given by  \cite{Wang2007b}:
\begin{eqnarray}
            \label{taupar1}\tau_{1,2}&=&2/(A\pm\sqrt{A^2-4B})\\[0.01 \textwidth]
            \label{taupar2}A&=&k_{12}+k_{21}+k_{23}+k_{31}\\[0.01 \textwidth]
            \label{taupar3}B&=&k_{12}k_{23}+k_{12}k_{31}+k_{21}k_{31}+k_{23}k_{31}\\[0.01 \textwidth]
            \label{apar} a&=&\frac{1-\tau_2k_{31}}{k_{31}(\tau_2-\tau_1)}
\end{eqnarray}
The parameter $\tau_1$ governs the antibunching, while $\tau_2$ governs the bunching of the $g^{(2)}$ function. The parameter $a$ determines how pronounced the bunching is. In contrast to Eq.\ (\ref{g23level}), the measured $g^{(2)}$ functions display $g^{(2)}(0)\ne 0$. For several emitters, this deviation is only due to the instrument response of the Hanbury Brown Twiss setup, i.e., in particular the timing jitter of the APDs (details see \cite{Neu2011}): Eq.\ (\ref{g23level}) convoluted with the instrument response fully explains the measured data [see Fig.\ \ref{fig:g2examples}(a)+(c)]. For emitters ND3 and NI1, the deviation $\Delta g^{(2)}(0)$ between the fitted value of $g^{(2)}(0)$ and the measured datapoints is less than 0.05, witnessing very pure single photon emission with negligible background contribution. For other emitters, broadband background emission of the diamond material deteriorates the $g^{(2)}$ functions. For the spectral region of interest, the broad luminescence is attributed to $sp^2$ bonded disordered carbon (in diamond films) introducing electronic states into the bandgap (e.g., \cite{Bergman1994}) or to grain boundaries in the diamond material \cite{Mora2003}. To take into account background luminescence, we follow Ref.\ \cite{Brouri2000} and include the probability $p_e$ that a detected photon stems from the single SiV center into the fit of the measured correlation function $g^{(2)}_m(\tau)$ via
\begin{equation}
g^{(2)}_m(\tau)=1+(g^{(2)}(\tau)-1)p_e^2.
\label{g2background}
\end{equation} $p_e$ is obtained from the signal to background ratio in the saturation curves.
From the fits of the $g^{(2)}$ function, we obtain the excitation power dependent values of the parameters $a$, $\tau_1$ and $\tau_2$ [see Eq.\ (\ref{g23level})]. In the following, we aim at modeling the power dependence of these parameters and deduce the rate coefficients of the color center's level scheme. Examples of the measured power dependent parameters for six individual SiV centers are given in Fig.\ \ref{g2par}.

\label{pumpdiscussion}In a first approach, we assume the rate coefficients $k_{21}$, $k_{23}$, $k_{31}$ to be constant, whereas $k_{12}$  depends linearly on the excitation power $P$: $k_{12}=\sigma P$. The assumption is justified, as the color center is excited to vibrationally excited states that typically relax within picoseconds to the vibrational ground state in the excited state (state 2) \cite{Lounis2005}. Thus, the intermediate pumping levels do not accumulate population, therefore efficiently suppressing stimulated emission and thus saturation on the pumping transition.  Under this assumption, $k_{21}$, $k_{23}$, $k_{31}$ can be calculated from high and vanishing excitation power limiting values of $a$, $\tau_1$ and $\tau_2$ (details see \cite{Neu2011}), whereas
 $\sigma$ can be derived using the saturation power $P_{sat}$
\begin{equation}
\sigma=\frac{k_{23}k_{31}+k_{21}k_{31}}{(k_{23}+k_{31})P_{sat}}. \label{Eqsigma}
\end{equation}
As visible from Fig. \ref{g2par}, the observed power dependence of $a$, $\tau_1$ and $\tau_2$ allows to estimate the high and low power limiting values of these parameters. Using the model with constant rate coefficients, we calculate the power dependent curves for $a$, $\tau_1$ and $\tau_2$, shown in Fig. \ref{g2par} as red dashed lines.
As evident from Fig. \ref{g2par}, the model with constant rate coefficients reasonably well describes the power dependence of $\tau_1$, and, for some emitters, the power dependence of $a$. Nevertheless, it totally fails explaining the power dependence of $\tau_2$: $\tau_2$ significantly increases up to three orders of magnitude at low excitation power. In contrast, the model predicts a nearly constant value of $\tau_2$ at low excitation power.

In Ref.\ \cite{Neu2011}, we tentatively suggested an extension of the simple three level model discussed above allowing to account for these deviations. We here verify this model using a larger number of emitters together with extended $g^{(2)}$ measurements.  To extend the model, an additional intensity dependent transition process is included. Following approaches in the literature \cite{Aharonovich2010a,Fleury2000}, we assume that the process reactivating the color center from the shelving state (de-shelving process, rate coefficient $k_{31}$) is intensity dependent. In order to accurately describe our experimental results, however, we find that the simple linear excitation power dependence of  $k_{31}$ found in \cite{Aharonovich2010a,Fleury2000} has to be replaced by a saturation behavior:
\begin{equation}
k_{31}=\frac{d\cdot P}{P+c}+k_{31}^0, \label{satdeshrate}
\end{equation}
where $k_{31}^0$ is an intensity independent part. As depicted in Fig.\ \ref{3levelschemedesh}, the de-shelving might be realized via an excitation from the shelving state to higher lying states that returns the color center to the ground state (see \cite{Fleury2000}). Such an excitation process might intrinsically exhibit a saturation behavior.

For this new model, we calculate $k_{23}$, $k_{21}$, $k_{31}^0$, $d$ under the assumption $k_{21}+ k_{23}> k_{31}^0$:
\begin{eqnarray}
\label{limnew1} k_{31}^0=&\frac{1}{\tau_2^0}\\
\label{limnew2}d=&\frac{\frac{1}{\tau_2^{\infty}}-(1+a^{\infty})\frac{1}{\tau_2^0}}{a^{\infty}+1}\\
\label{limnew3}k_{23}=&\frac{1}{\tau_2^{\infty}}-k_{31}^0-d\\
\label{limnew4} k_{21}=&\frac{1}{\tau_1^0}-k_{23}
\end{eqnarray}
The superscript ${}^{\infty}$ (${}^{0}$) denotes the limit for high (vanishing) excitation power. To derive the power dependence of $a$, $\tau_1$ and $\tau_2$, two additional parameters have to be determined: $c$ [see Eq.\ (\ref{satdeshrate})] and  $\sigma$. We can no longer obtain $\sigma$ from the saturation curve as it is no longer feasible to link the rate coefficients to  $P_{sat}$ as in Eq.\ (\ref{Eqsigma}). Instead, $c$ and $\sigma$ have to be obtained from fits of the power dependent curves of $a$, $\tau_1$ and $\tau_2$: We employ Eqs.\  (\ref{taupar1})--(\ref{apar}), together with the definitions of $k_{12}$ and $k_{31}$ [Eq.\ (\ref{satdeshrate})], as fit functions with free parameters $c$ and $\sigma$.  From the fits, we find that the power dependence of $\tau_1$ is almost fully determined by $\sigma$. Thus, we determine $\sigma$ from the fit of $\tau_1(P)$. Using the resulting value of $\sigma$ as fixed parameter, we fit $a(P)$ and obtain the value of $c$. To complete the description, we plot the resulting curve for $\tau_2(P)$ with these parameters. Using this procedure, we find the best accordance of the fits, displayed in Fig.\ \ref{g2par} as solid blue lines, and the measured data.  Table \ref{tab_deshelvrates} summarizes the rate coefficients and fit parameters obtained for the individual SiV centers covered in Fig.\ \ref{g2par}.
\begin{table}
\centering
\caption{Rate coefficients deduced from the limiting values of $a$, $\tau_1$ and $\tau_2$ using the three level model including intensity dependent de-shelving and parameters $c$ and $\sigma$ obtained from the fits. For comparison also $P_{sat}$ is given. \label{tab_deshelvrates}}
\begin{tabular}{cccccccc}
\hline
  & $k_{21}$(MHz) & $k_{23}$(MHz)& $k_{31}^0$(MHz) &d (MHz) &$\sigma$(MHz/$\mu$W)& $c$ ($\mu$W)& $P_{sat}$ ($\mu$W) \\
\hline
ND1 & 4408  & 137.0 & 0.27& 18.6  & 12.0  & 11.9 &30.6 \\
ND2 & 3424  & 24.6 & 1.7& 24.4 & 8.9 & 177&167  \\
ND3 & 771  & 23.3 & 0.35& 24.7 & 5.7 & 57 &105.3 \\
ND4 & 1084 & 31.7 & 0.12& 13.1 & 7.0 & 2743&282  \\
NI1 & 3479 & 92.6 & 0.82& 45.5 & 4.2 & 1067 &692 \\
NI7 & 1638 & 1.5 & 0.16& 0.7 & 7.2 & 300&46.9 \\
\hline
\end{tabular}
\end{table}

As apparent from Fig.\ \ref{g2par}, using the extended model we obtain a much better concurrence of the fitted curves and the measured power dependence of $a$, $\tau_1$ and $\tau_2$: Especially for emitter NI7 [Fig.\ \ref{g2par}(f)], ND2 [Fig.\ \ref{g2par}(b)] and ND3 [Fig.\ \ref{g2par}(c)] all curves are well described. For emitters ND4 [Fig.\ \ref{g2par}(d)] and ND1 [Fig.\ \ref{g2par}(a)], the rapid drop of $\tau_2$ at intermediate powers is overestimated. This has been observed for several other emitters. Additionally, the power dependence of $a$ can only be qualitatively  described using the model with intensity dependent de-shelving for emitter ND4 [Fig.\ \ref{g2par}(d)]. For NI1 [Fig.\ \ref{g2par}(e)], an extraordinary behavior of $a$ is observed, including an increase of $a$ at very low powers. Note that NI1 is the brightest color center observed delivering $I_\infty=6.2$ Mcps.

In the following, we shortly summarize the sources of errors in the data evaluation and their consequences for the obtained rate coefficients $k_{ij}$. First, $k_{31}^0$ comprises a comparably large uncertainty as this rate coefficient is determined by $\tau_2^0$ [see Eq.\ (\ref{limnew1})]: The estimation of $\tau_2^0$ from the very steep curves at low excitation powers is challenging. Furthermore, $a$ is often rather small at low excitation power (weak bunching), thus the proper determination of $\tau_2$ is demanding.  Second, $k_{21}$, in contrast, can be reliably assigned as it is determined by $\tau_1^0$ [see Eq.\ (\ref{limnew4})]: $\tau_1^0$ can be  precisely estimated as $\tau_1(P)$ has a comparably small slope for low excitation powers. Third, $k_{23}$ includes a moderate uncertainty as it is governed by $\tau_2^{\infty}$ and $a^{\infty}$ [Eq.\ (\ref{limnew3})]:  Fig.\ \ref{g2par}(a) illustrates the challenge in finding $a^{\infty}$. The power dependence of $a$ includes two datasets: The one marked with filled squares (filled dots) has been obtained including (excluding) background correction for the $g^{(2)}$ function fitting. Both fits well describe the measured $g^{(2)}$ function and yield very similar values for $\tau_1$ and $\tau_2$, respectively, but differing values for $a$: The instrument response washes out the $g^{(2)}$ function, thus a steep slope of $g^{(2)}$ in combination with a high absolute value close to zero delay leads to an increase of $g^{(2)}(0)$ similar to the modification owing to background fluorescence.
Thus, an uncertainty in the background correction induces an uncertainty in $a^{\infty}$.  In contrast, $\tau_2^{\infty}$ can be extracted reliably from the measured data as a clear convergence toward a constant value is observed for most emitters.
\begin{figure}[h]
\begin{center}
\includegraphics[width=0.9\textwidth]{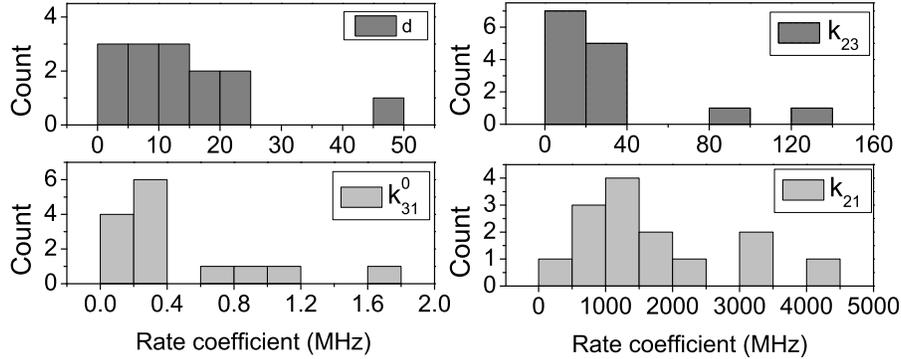}
\caption{Histograms of rate coefficients $k_{21}$, $k_{23}$, $k_{31}$ and $d$ obtained from the model of saturating de-shelving. The histograms take into account emitters from NIs as well as randomly oriented NDs.  \label{g2hist}}
\end{center}
\end{figure}

In summary, the $g^{(2)}$ measurements unambiguously reveal the presence of a shelving state and verify the existence of an intensity dependent de-shelving path. However, at present we cannot identify the nature of the shelving state and its energetic position nor the transition responsible for the de-shelving.
The rate coefficients obtained from the intensity dependent de-shelving model for 14 emitters are summarized in Fig.\ \ref{g2hist}. It is clear that, for all emitters, $k_{21}$ is significantly higher than the other rate coefficients. The significant spread of $k_{21}$ might be due to the local environment as well as a varying quantum efficiency of the transition (for  a detailed discussion see Sec.\ \ref{sec_quantumeff_colleff}). $k_{31}^0$ is lower than 1 MHz for the majority of the emitters, indicating a long lived shelving state. $d$, representing the high power limit of the de-shelving rate coefficient, is at least a factor of 4.6, for most emitters even an order of magnitude, larger than $k_{31}^0$. Comparing the parameter $c$, indicating the saturation power for the de-shelving process, with the measured saturation power $P_{sat}$, we find that for several emitters $c$ and $P_{sat}$ have a similar value. The values are summarized in Tab.\ \ref{tab_deshelvrates}. One thus might suspect, that the saturation of the de-shelving transition determines the saturation of the fluorescence of the SiV center. The parameter $\sigma$ gives the absorption cross section of the SiV centers (for details on the conversion from excitation power to intensity see Sec.\ \ref{sec_satroomtemp}). The values of $\sigma$ given in Tab.\ \ref{tab_deshelvrates} correspond to absorption cross sections of 1.4--4.2$\times$10$^{-14}$ cm$^2$. Note that the excitation has been performed with optimized linear polarization to address the single transition dipole moment of the SiV center. The absorption cross section for the nitrogen vacancy (NV) center under 532 nm excitation, averaged over the possible orientations of the transition dipole moments, has been recently determined to be approx. 1$\times$10$^{-16}$ cm$^2$ \cite{Chapman2011}. The absorption cross section for the NE8 center, a nickel-nitrogen complex, has been determined as 1.7$\times$10$^{-16}$ cm$^2$ in \cite{Wu2006a} for 687 nm excitation. The values for $\sigma$ determined for the SiV center here thus exceed the absorption cross section of the NV center by two orders of magnitude. To further clarify this issue, further investigations, e.g.,  using pulsed laser excitation as in \cite{Chapman2011} are desirable to determine the absorption cross section independently.  For  $k_{23}$, a large spread is observed ranging from 137 MHz to 1.5 MHz, thus we conclude that the coupling to the shelving state strongly varies among different emitters. It is also apparent from Fig.\ \ref{g2hist} as well as Tab.\ \ref{tab_deshelvrates} that $k_{23}$ is comparable to or even larger than $k_{31}$, even at high excitation powers. Thus, population accumulates in the shelving state. The influence of the shelving state will be further addressed in Sec.\ \ref{sec_quantumeff_colleff}.
\section{Photostability of single SiV centers \label{photostability}}
\begin{figure}[h]
\centering
\includegraphics[width=\textwidth]{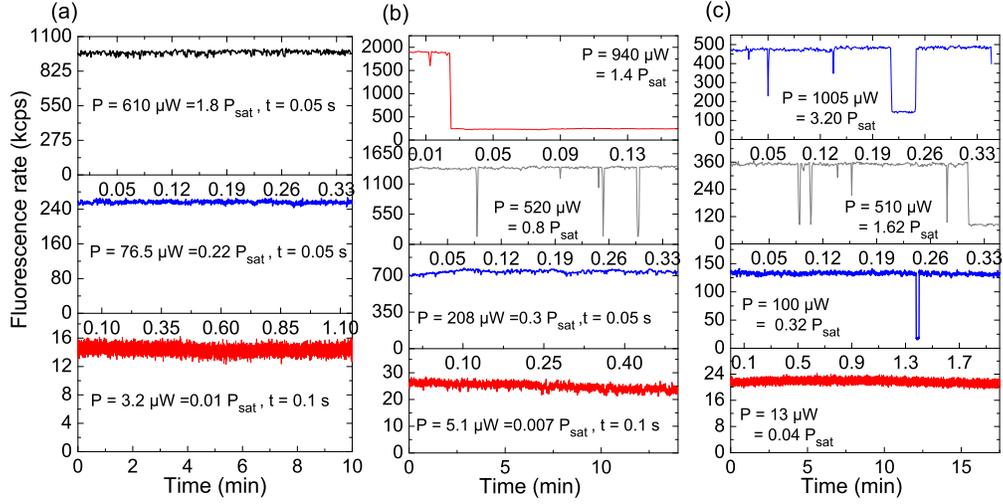}
\caption{Fluorescence timetraces of single SiV centers (a) SiV center with fully photostable emission (emitter ND4, excited at 695 nm). (b) Emitter with destabilization at higher excitation power and permanent photobleaching (emitter NI1). (c) Emitter with longer time intervals of fluorescence intermittence (emitter NI6). The count rate of each emitter has been calculated in time windows of 100 ms for the lowest excitation power (50 ms for higher excitation powers). \label{timetraces}}
\end{figure}

For single photon generation using optical excitation, the photostability of the emitter is crucial: Permanent or temporal loss of single photon emission, i.e., photobleaching or blinking, under optical excitation limits the applicability of the single photon source. For single color centers in diamond, photobleaching has been reported in the literature for single centers emitting in the near-infrared spectral region \cite{Siyushev2009}, for single NV centers in NDs \cite{Bradac2010} and for a center emitting at 736.8 nm \cite{Wang2007b}. However, none of the publications discusses the origin of the permanent bleaching.
In addition, fluorescence intermittence (blinking) is possible.  For single color centers in NDs, blinking has been observed in Ref.\ \cite{Bradac2010} and is attributed to trapping and release of charges on the surfaces of the NDs.
To analyze the photostability of single SiV centers, we obtain time traces of the fluorescence rate (see Fig. \ref{timetraces}).
Based on the observed fluorescence stability, we can very roughly arrange the observed emitters into three classes as discussed below, whereas emitters may belong to class 2 and 3 simultaneously.
\subsection*{Class 1: emitters with fully photostable emission} An example for the fluorescence time trace of a photostable emitter (ND4) is given in Fig.\ \ref{timetraces}(a). These emitters are photostable for excitation powers far above saturation: E.g., emitter ND1 has been shown to be stable under excitation powers up to $32P_{sat}$. The typical observation time for the intensity dependent $g^{(2)}$ measurements discussed above is about one hour, thus these emitters have shown to be photostable for at least one hour under continuous laser excitation. Roughly 20--30\% of the emitters investigated in detail show full photostability. As visible, e.g., in the lowermost graph of Fig.\ \ref{timetraces}(b), the count rate of the stable emitters nevertheless exhibits slow variations (timescale $>$ 5 min) due to spatial drifts of the emitter out of the laser focus which can be undone repositioning the emitter.
\subsection*{Class 2: emitters exhibiting fluorescence intermittence}
Figures \ref{timetraces}(b) and (c) give time traces for emitters exhibiting only partially stable emission due to  fluorescence intermittence. We find dark times ranging from several 100 ms up to 2 min. Figure \ref{timetraces}(b) and (c) indicate a general trend: The probability for blinking events is apparently higher at elevated excitation power, thus one might suspect that the transition to the dark state is induced by the pump light. For excitation powers below or close to saturation, also for these partially stable emitters almost constant single photon emission can be obtained as the blinking events are rare.
We point out that in Ref.\ \cite{Wang2007b} an individual blinking color center with a ZPL wavelength compatible with SiV emission wavelengths \cite{Neu2011b} is reported, evidencing the blinking of SiV centers in single crystal bulk diamond.
\subsection*{Class 3: emitters for which permanent photobleaching occurred}
Figure \ref{timetraces}(b) shows an emitter for which permanent photobleaching occurred at elevated power. 'Permanent' here means, in general, that for waiting times of at least 10 min (partially without laser illumination) no recovery of the fluorescence has been detected.
For emitter NI1 [Fig.\ \ref{timetraces}(b)], prior to the permanent bleaching event, blinking was observed with a trend to enhanced blinking activity for higher laser powers. However, we also found emitters that were bleached without any prior sign of fluorescence instability/intermittence preferably at higher excitation powers and after longer observations times (e.g., emitter NI3 bleached at $13P_{sat}$, observation time 1 hour). We point out that in Ref.\ \cite{Wang2007b}, permanent bleaching of the investigated blinking center after one week of observation is reported.
\subsection{Discussion of the observations}
Blinking of color centers can be due to photoionization of the color centers as the charge state after ionization can be non-radiative or emits at a wavelength beyond the preselected spectral window (here 730--750 nm). We here are not able to verify this option as the detection efficiency of our setup for the weak luminescence at 946 nm found to originate from an alternative charge state of the SiV center \cite{Dhaenens2011} is too low. For NV centers in NDs with 5 nm size, blinking has been interpreted in terms of the capture of electrons in surface traps \cite{Bradac2010}. However, simultaneously it was found that NV$^-$ to  NV$^0$ conversion is not responsible for blinking or bleaching as no  NV$^0$ luminescence was observed \cite{Bradac2010}. The authors of Ref.\ \cite{Bradac2010} use the analogy of the optical excitation to an exciton formation to explain this behavior: As long as the electron of the exciton is captured, no fluorescence occurs. Other authors, in contrast, suspect that the lack of excess electrons needed to charge/recharge NV centers in small NDs leads to photobleaching after photoionization \cite{Tisler2009}.
The observation of photostable SiV centers shows that the SiV complex is, in principle, photostable under red laser excitation. thus, modifications of the color center's local environment have to induce blinking or bleaching. For the SiV centers, the corresponding mechanisms are not clear. However, due to the enhanced probability for the centers to undergo blinking transitions at elevated excitation powers, we suggest that the transition to the dark state is photoinduced as also found for NV centers in \cite{Bradac2010}. It might also be possible that the color center undergoes spontaneous transitions from its excited state to the dark state. Thus, with a higher excited state population the rate for a transition to the dark state is enhanced.

We here choose the photon energy of the excitation laser sufficiently low so that the (spatially localized) electrons bound to the SiV center cannot be excited from the ground state of the SiV center to (spatially delocalized) conduction band states. Delocalized states may promote the capture of electrons by traps in the vicinity of the color centers: An electron in such delocalized states has a finite probability to be found at the spatial site of a trap. In addition to the primary excitation process, one might think of further excitation of the bound electrons from the excited state of the color center (excited state absorption). Also a simultaneous absorption of two photons is a possible route to photoionization (c.f. two photon absorption processes for NV centers see, e.g., \cite{Wee2007}). In both cases, trapping of electrons promoted to the conduction band could subsequently induce fluorescence intermittence. The nature of the trapping states is not clear for the SiV centers investigated here. One might think of surface states as, e.g., in Ref.\ \cite{Bradac2010}; however, as we use CVD diamonds, the surfaces of the diamonds should contain significantly less graphite and disordered carbon as compared to the surfaces of the detonation NDs used in \cite{Bradac2010}. A second possibility might be trapping of the electrons at other impurity atoms, e.g., substitutional nitrogen atoms \cite{Ulbricht2011}.

Additionally, it is not clear whether the mechanisms responsible for blinking and permanent photobleaching are identical. As very long blinking times have been observed, it is possible that the blinking mechanism is also responsible for the 'permanent' bleaching and that a recovery of the fluorescence after long waiting times is possible but has not been observed. This might especially happen if the laser is not able to free the electrons from their trapping states and if the spontaneous, possibly thermal, escape from these traps is very unlikely. The latter argument suggests trapping states deep within the band gap.
The observation of photostable SiV centers is very promising for the application of single SiV centers as single photon sources. Using surface treatments as well as further control of the impurity content might help to enhance the fraction of fully photostable SiV centers.
\section{Quantum efficiency of single SiV centers} \label{sec_quantumeff_colleff}
This section deals with the maximum single photon rates $I_{\infty}$ observed for single SiV centers and the effects limiting this rate. In particular, we aim at deducing the quantum efficiency $\eta_{qe}$ of the SiV centers.
$I_{\infty}$ for continuous laser excitation is given by:
\begin{equation}
I_{\infty}=\eta_{det}\,\eta_{qe}\, k_{21}\, N_2^{\infty}(P\rightarrow \infty)=\eta_{det}\, \eta_{qe}\,\frac{k_{21}}{1+\frac{k_{23}}{k_{31}^0+d}} \label{glqe}
\end{equation}
$ N_2^{\infty}(P\rightarrow \infty)$ is the maximum steady state population of the excited state.
$k_{21}$, $k_{23}$, $k_{31}^0$ and $d$ are the rate coefficients obtained from the intensity dependent de-shelving model. $\eta_{det}$ is the detection efficiency of the experimental setup. It is the product of the collection efficiency $\eta_{coll}$, i.e., the probability to collect an emitted fluorescence photon, and the internal efficiency of the detection setup $\eta_{det}^{int}$, i.e., the probability to detect a collected photon. Taking into account the transmission/reflection of all optical components, as well as the APD detection efficiency and the efficiency of the multi-mode fiber coupling in the employed confocal microscope setup (details see \cite{Neu2011,Neu2011b}), we estimate the internal detection efficiency $\eta_{det}^{int}$ of our setup as 25\%.  $\eta_{qe}$ is the quantum efficiency of the SiV center, i.e., the probability for a photon emission upon a transition from state 2 to 1 (see Fig.\ \ref{3levelschemedesh}).  First, we determine the influence of the shelving state on $I_{\infty}$. For an off-resonantly pumped two level system, assuming a very fast relaxation to state 2 after excitation, full population inversion $ N_2^{\infty}(P\rightarrow \infty)=1$ can be obtained. For the emitters discussed here, we obtain $ N_2^{\infty}(P\rightarrow \infty)$ as summarized in Tab.\ \ref{tab_n2population}. As apparent from Tab.\ \ref{tab_n2population},  the influence of the shelving state on $I_{\infty}$ differs for individual emitters: For emitter ND3, $ N_2^{\infty}(P\rightarrow \infty)$ is only lowered by a factor of two compared to the two level case. On the other hand, for emitter ND1, $ N_2^{\infty}(P\rightarrow \infty)$ is smaller by nearly an order of magnitude compared to the off-resonantly pumped two level system. As apparent from Tab.\ \ref{tab_n2population}, the shelving rate $k_{23}$ is always much smaller compared to $k_{21}$. However, due to a slow depopulation of the shelving state, for several SiV centers, the shelving state accumulates most of the population, leading to a significant loss of brightness compared to a two level system.
\begin{table}
\centering
\caption{Rate coefficients $k_{ij}$, maximum excited state population $N_2^{\infty}(P\rightarrow \infty)$, maximum photon rate $I_{\infty}$ and quantum efficiency $\eta_{qe}$ for individual SiV centers. To calculate the quantum efficiency, we use a collection efficiency of 78\% (28\%) for a parallel (perpendicular) dipole, corresponding to an emitter distance of 75 nm.    \label{tab_n2population}}
\begin{tabular}{p{0.7cm}p{1.0cm}p{1.0cm}p{1.0cm}p{0.9cm}p{1.8cm}p{1.0cm}p{0.7cm}p{0.7cm}}
\hline
  &  $k_{21}$ (MHz) & $k_{23}$ (MHz)& $k_{31}^0$ (MHz) &d (MHz) & $N_2^{\infty}$ $(P\to \infty)$& $I_{\infty}$ (Mcps) & $\eta_{qe}^{\parallel}$ (\%) & $\eta_{qe}^{\perp}$ (\%)\\
\hline
ND1	&4408	&137&	0.27&	18.6&	0.12&	0.84&	0.8&	2.2\\
ND2&	3424&	24.6&	1.7	&24.4&	0.51&	1.53&	0.4	&1.2\\
ND3	&771&	23.6&	0.35&	24.7&	0.51&	2.46&	3.2&	8.9\\
ND4&	1084&	31.7&	0.12&	13.1&	0.29&	2.06&	3.3&	9.2\\
ND5	&1545.1&	17.4&	1	&11.9&	0.43	&2.39&	1.9&	5.2\\
ND6	&770.1&	11.1&	0.79&	5.65&	0.37&	0.78&	1.4&	3.9\\
ND7	&1053.6&	21.7&	0.11&	3.44&	0.14&	0.59&	2.1	&5.7\\
NI1 &	3479&	92.6&	0.82&	45.5&	0.33&	6.24&	2.8&	7.7\\
NI3	&161&	7.3&	0.24&	11.9&	0.62&	0.17&	0.9&	2.4\\
NI7 &	1638&	1.5&	0.16&	0.7	&0.36	&0.34&	0.3&	0.8\\
NI8	&2487&	12.5&	0.15&	5.3&	0.30&	0.9&	0.6&	1.7\\
NI9&	1181.7&	1.8&	0.21&	3.1&	0.65&	3.82&	2.6&	7.1\\
NI10	&798.8&	34.6&	0.22&	16.2&	0.32&	0.8	&1.6&	4.4\\
NI11	&1076&	13.3&	0.32&	8.2&	0.39	&0.52&	0.6&	1.8\\
\hline
\end{tabular}
\end{table}
\subsection{Calculation of collection efficiency and quantum efficiency: influence of the Ir substrate}
The radiation properties of SiV color centers in NDs/NIs on Ir
are investigated by considering a point-like oscillating dipole near a metal surface~\cite{chance78}. The orientation of the individual color center dipoles is unknown, we thus investigate the limiting cases of a dipole perpendicular and parallel to the interface in our simulations. Determining the exact position of the emitting SiV centers and thus their distance from the Ir surface is, in principle, not possible as they are created at an unknown instant of time during the CVD growth of the nanodiamonds, i.e., at an unknown position inside the CVD nanocrystal. For an estimation of the characteristic distance of an emitter from the surface, we assume the emitter to be located roughly in the center of the NDs/NIs, i.e., approx. 40--100 nm above the metal film. These assumptions
seem to be crude approximations as we are neglecting the fact that the color center
is inside a dielectric nanoparticle, which may affect the photophysics of the SiV center through
changes in the spontaneous emission rate~\cite{rogobete03}
and different regimes of interaction with the metal surface~\cite{chen12}.
Nonetheless, since we are interested in a qualitative understanding of the role of the Ir surface and we are concerned with distances above roughly 50 nm,
it turns out that our approach is sufficiently sophisticated to describe the most
important involved phenomena.

\begin{figure}[htpb]
\centering
\includegraphics[width=\textwidth]{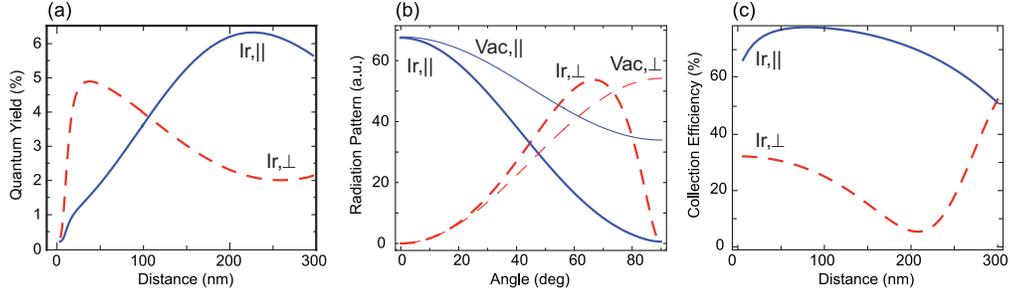}
\caption{Emissions characteristics of a point-like oscillating dipole in vacuum near an Ir surface for orientation parallel (blue solid curves) and perpendicular (red dashed curves) to the interface.
The emission wavelength is 740 nm and the intrinsic quantum yield is 5\%.
(a) Effective quantum yield as a function of distance from the Ir surface.
(b) Radiation pattern for a dipole located 80 nm above the Ir surface.
The thin curves refer to a dipole in free space.
(c) Collection efficiency as a function of distance for a microscope objective with NA=0.8.
\label{dipolemirror}}
\end{figure}

The dipole emits radiation in vacuo at $\lambda=740$ nm, where the dielectric function of Ir takes
the value $\epsilon_\mathrm{Ir}=-18+25\mathrm{i}$~\cite{palik98}.
In practice, the Ir surface modifies the quantum yield by changing the spontaneous
emission rate and absorbing a fraction of the emitted light.
If $\gamma_\mathrm{0}$ and $\eta_0$ are the intrinsic radiative decay rate and
quantum yield of the SiV center respectively, an expression for the effective quantum yield
reads
\begin{equation}
\eta=\frac{\eta_0}{(1-\eta_0)\gamma_0/\gamma_\mathrm{r}+\eta_0/\eta_\mathrm{a}}.
\end{equation}
$\gamma_\mathrm{r}$ represents the modified radiative decay rate and $\eta_\mathrm{a}$
accounts for the rate $\gamma_\mathrm{nr}$ of energy dissipation in the metal,  $\eta_\mathrm{a}=\gamma_\mathrm{r}/(\gamma_\mathrm{r}+\gamma_\mathrm{nr})$.
The effective quantum yield for a parallel and a perpendicular dipole is shown in Fig.~\ref{dipolemirror}(a)
for $\eta_0=5\%$ (according to \cite{Turukhin1996}) as a function of distance from the Ir surface.
Note that the competition between $\eta_\mathrm{a}$ and $\gamma_\mathrm{r}/\gamma_0$
may lead to an effective quantum yield larger than $\eta_0$ at certain
distances if the dipole is parallel to the Ir surface. The effect is, however, modest
being $\gamma_\mathrm{r}$ at most a factor of two larger than $\gamma_0$.
For very short distances, the increase of $\gamma_\mathrm{nr}$ for both dipole
orientations due to near-field energy transfer, gives rise to the well-known phenomenon of
fluorescence quenching~\cite{chance78}.

The collection efficiency is given by the fraction of power radiated in the solid angle
determined by the numerical aperture (NA) of the microscope objective, divided
by the total radiated power which, in the presence of a metal surface, is
limited to the upper half space.
In short, the calculations are performed by expanding the dipole field
in plane waves. Each partial wave fulfills the boundary conditions at the interface
through Fresnel coefficients. Further details can be found
in~\cite{hellen87,novotny06}.
The collection efficiency strongly depends on the radiation pattern, which is
significantly modified by the Ir surface. An example for parallel and perpendicular dipoles
located 80 nm above the metal is shown in Fig.~\ref{dipolemirror}(b), where
the pattern for the respective dipoles in free space is added for comparison.
In practice, the Ir surface channels the emission toward smaller angles, thus increasing the fraction of emission that can be collected by the microscope objective. By changing the distance,
the radiation pattern varies differently for the two relevant dipole
orientations, as it can be inferred from Fig.~\ref{dipolemirror}(c), where the
collection efficiency using a microscope objective with NA=0.8 is plotted as
a function of distance. It is found that a parallel dipole is a favorable
configuration for obtaining large count rates, where the fraction
of collected photons can exceed 70\% with
standard optics for microscopy.
We recall that for the case of SiV centers in bulk the large refractive index of diamond
leads to poor collection efficiencies of only up to a few percent,
for a dipole parallel to the interface (not shown). For the typical emitter distances of 40-100 nm, we find a collection efficiency of approx. 75\% (30\%) for parallel (perpendicular) dipoles, each value varying within less than 10\% [see Fig.~\ref{dipolemirror}(c)].

In the following, we compare our experiments using NDs/NIs on Ir to previous experiments using SiV centers in bulk diamond in \cite{Wang2006,Wang2007b}. For a parallel dipole, the collection efficiency in the NDs/NIs on Ir is enhanced by approx.\ a factor of 20 compared to SiV centers in bulk diamond. On the other hand, the maximum photon rates $I_{\infty}$ we find are about three orders of magnitude higher than observed in \cite{Wang2006,Wang2007b}. $\eta_{det}^{int}$ in Ref.\ \cite{Wang2007b} is 19\% and thus close to the value of our setup (25\%). However, it should be noted that \cite{Wang2007b} does not include a discussion of the transition rate coefficients similar to the discussion given here. The quantum efficiency of a single SiV center modeled as a two level system is estimated to be 0.5\% in \cite{Wang2007b}.  Therefore, we tentatively suggest that the high brightness of some of the SiV centers in CVD NDs/NIs cannot fully be attributed to an enhanced collection efficiency but is also correlated to a slightly higher quantum efficiency for the \textit{in situ} produced SiV centers as discussed below.
\subsection{Experimental estimation of the quantum efficiency}
We now use $\eta_{coll}$ calculated for a dipole located 75 nm above the Ir surface to estimate $\eta_{qe}$ of the SiV centers according to Eq.\ (\ref{glqe}). We find values ranging from $\eta_{qe}^{\parallel}=0.3\%$ to $\eta_{qe}^{\perp}= 9.2\%$ assuming the two limiting cases of parallel and perpendicular dipole orientations (Tab.\ \ref{tab_n2population}). Thus, the observed quantum efficiency is comparable to previous measurements on SiV ensembles in polycrystalline films yielding $\eta_{qe}=5\%$ \cite{Turukhin1996}. The values determined here, however, do not straightforwardly  represent the internal quantum efficiency, i.e., the probability for a radiative decay of the SiV center in bulk diamond. First, for the saturation measurements, from which we obtain $I_{\infty}$, only the fluorescence in a spectral window 730--750 nm is recorded, thus neglecting roughly 30\% of the fluorescence emitted into sidebands and additional electronic transitions in the near infrared spectral range (see \cite{Neu2012}). Thus, these transitions are erroneously considered as non-radiative and the quantum efficiency $\eta_{qe}$ will be underestimated by approx. 30\%. However, as the fractional intensity of the red-shifted emission significantly varies for individual emitters \cite{Neu2011b} also the error for the estimation of $\eta_{qe}$ varies.  In addition to influencing $\eta_{coll}$, the presence of the metal can also quench the fluorescence for emitters close to the surface, thus reducing $\eta_{qe}$.  However, to estimate this influence, the unknown intrinsic quantum yield of the SiV centers as well as the exact distance to the metal has to be taken into account. Thus, $\eta_{qe}$ estimated here is the probability for a photon emission in a restricted spectral range for an SiV center above the metal surface. Furthermore, in addition to our simple model of a dipole in air above the metal surface, the ND can strongly modify the radiation pattern of the emitting dipole: As discussed in Refs.\ \cite{Greffet2011,Castelletto2011} and mentioned before, in spherical particles with sizes comparable to the wavelength of the fluorescence and excitation light, resonant modes (Mie-resonances) can develop, strongly modifying  the radiation pattern and thus $\eta_{coll}$: Ref.\ \cite{Castelletto2011} demonstrates a variation of $\eta_{coll}$ between approx.\ 1\% and 20\% (spherical NDs on a sapphire substrate, size varies from 50--200 nm). However, spherical NDs still simplify the problem as the nanocrystals mostly resemble cubo-octahedral shapes and their exact size is unknown \cite{Neu2011,Neu2011b}. We cannot estimate the modification of $\eta_{coll}$ due to possible resonant modes in our NDs/NIs of unknown shape and size. Therefore, we cannot state whether this effect leads to an overestimation or underestimation of $\eta_{qe}$. In addition to these considerations, we emphasize that the rate coefficients and thus the excited state population include uncertainties (discussed above). Furthermore, the measurement of the maximum obtainable photon rates includes an uncertainty due to the contribution of background fluorescence.
A quantum efficiency strongly deviating from 100\%, in principle, might have several reasons:
First, quenching of color center luminescence by the proximity to defect rich crystal areas has been reported in the literature: Refs.\ \cite{Smith2010}, \cite{Castelletto2011} and \cite{Grudinkin2012} indicate quenching due to graphite and disordered carbon on ND surfaces, crystal damage as a consequence of heavy ion irradiation and structural defects or non-diamond carbon phases. The actual process leading to the quenching is not discussed or identified. Despite a high crystal quality of the NDs/NIs employed in this work, defects like dislocations are present and might possibly induce a quenching.
Moreover, in a solid state host radiative transitions can be quenched by the direct emission of phonons (multi-phonon relaxation). The 1.68 eV transition of the SiV center equals 10.2 phonon energies (165 meV in diamond). Ref.\ \cite{Rogers2010} summarizes evidence for multi-phonon quenching for luminescent transitions in diamond with similar energy.
An additional quenching mechanism for color centers has been introduced by Dexter, Klick and Russel in \cite{Dexter1955}. The non-radiative process is induced at a crossing point in the configuration coordinate diagram where the energy of a low lying vibrational state in the electronically excited state matches the energy of a highly excited vibrational state in the ground state and the color center can relax to the ground state without emission of radiation. Such processes might exist for SiV centers in diamond but have not been considered so far.
\section{Conclusion}
Single SiV centers in NDs and NIs on Ir films have been shown to exhibit high brightness under continuous excitation. We have developed a model accurately describing the three level population dynamics of these centers including an intensity dependent de-shelving process. SiV centers have been observed to retain photostability for excitation well above saturation, however, also blinking or bleaching centers have been found. The employed material system, nanocrystals on Ir, enables a high fluorescence collection efficiency exceeding 70\%. With this observations, we estimate quantum efficiencies for single SiV centers of up to 9\%.
\section*{Acknowledgments}
We thank M. Fischer, S. Gsell and M. Schreck (University of Augsburg) for supplying the CVD diamond samples. The project was financially supported by the Bundesministerium f\"ur Bildung und Forschung within the projects EphQuaM (contract 01BL0903) and QuOReP (contract 01BQ1011). M. Agio wishes to thank F. Koenderink (AMOLF) and he acknowledges
financial support from the EU-STREP project ``QIBEC''.
\end{document}